\documentclass[aps,onecolumn,showpacs,nofootinbib]{revtex4}
\usepackage{graphicx}
\usepackage{epsfig}
\usepackage{amssymb}
\usepackage{graphicx}
\usepackage{amsfonts}
\usepackage{latexsym}
\usepackage{amsmath}
\usepackage{subfigure}
\usepackage{mathrsfs}
\numberwithin{equation}{section}

\newcommand\encadremath[1]{\vbox{\hrule\hbox{\vrule\kern8pt
\vbox{\kern8pt \hbox{$\displaystyle #1$}\kern8pt}
\kern8pt\vrule}\hrule}}
\def\enca#1{\vbox{\hrule\hbox{
\vrule\kern8pt\vbox{\kern8pt \hbox{$\displaystyle #1$} \kern8pt}
\kern8pt\vrule}\hrule}}

\newcommand{\bt}{(\beta)}
\def\prl#1#2#3{{ Phys. Rev. Lett.} {\bf #1}, #2 (#3)}

\def\pre#1#2#3{Phys. Rev. E  {#1} #2 #3}

\begin{document}
\title{Bulk asymptotics of skew-orthogonal polynomials for quartic double well potential
and universality in the matrix model}
\author{Saugata Ghosh}\email{saugata135@yahoo.com}
\affiliation{F-252 Sushant Lok II, Gurgaon 122002, India.}
\date{\today}
\begin{abstract}
We derive bulk asymptotics of skew-orthogonal
polynomials (sop) $\pi^{\bt}_{m}$, $\beta=1$, $4$, defined w.r.t. the weight
$\exp(-2NV(x))$, $V (x)=gx^4/4+tx^2/2$, $g>0$ and $t<0$. We assume that as $m,N \rightarrow\infty$
there exists an
$\epsilon > 0$, such that
$\epsilon\leq (m/N)\leq \lambda_{\rm cr}-\epsilon$, where $\lambda_{\rm cr}$ is the critical value which separates
sop with two cuts from those with one
cut. Simultaneously we derive asymptotics for the
recursive coefficients of skew-orthogonal polynomials. The proof
is based on obtaining a finite term recursion relation between  sop and  orthogonal polynomials (op) and using asymptotic results of op derived in \cite{bleher}.
Finally, we apply these asymptotic results of sop and their recursion coefficients
in the generalized Christoffel-Darboux formula (GCD) \cite{ghosh3}
to obtain level densities and sine-kernels in the bulk of the spectrum
for orthogonal and symplectic ensembles of random matrices.
\end{abstract}
\maketitle

\section{Introduction}

Skew-orthogonal polynomials are useful in the study of
orthogonal ($\beta=1$) and symplectic ($\beta=4$)
ensembles of random matrices \cite{dyson1,dysommehta,mehta,mehta1,ghosh,ghoshpandey,ghosh3,ghosh4,ghosh-classical,pandeyghosh,eynard,pierce}.
In this paper, we derive asymptotics of skew
orthogonal functions $\phi_{m}^{\bt}(x)$ and $\psi_{m}^{\bt}(x)$ and their recursion coefficients
\cite{ghosh3,ghosh4}, defined w.r.t. the weight
\begin{equation}
\label{quarticweight}
w(x)=\exp(-2NV(x)),\qquad V(x)=\frac{gx^4}{4}+\frac{tx^2}{2},\qquad g > 0,\qquad t<0.
\end{equation}
Here, $2N$ is a large parameter which, in the context of random matrix theory, is the
size of the matrices.

We define skew-orthogonal functions:
\begin{eqnarray}
\label{quasipolynomial} \phi_{n}^{\bt}(x) =
\frac{1}{\sqrt{g^{\bt}_n}}\pi^{\bt}_{n}(x)\exp(-NV(x)),\qquad
 \pi^{\bt}_{n}(x) =
\sum^{n}_{k=0}c^{(n,\beta)}_{k}x^k,\qquad \beta=1,4,\\
\label{psi} \psi^{(4)}_{n}(x) :=
  \phi '^{(4)}_{n}(x),\qquad
\psi^{(1)}_{n}(x):=\int_{\mathbb R}\phi^{(1)}_{n}(y)\epsilon(x-y)dy,\qquad \epsilon (r)=\frac{|r|}{2r},
\qquad n\in\mathbb N,
\end{eqnarray}
where $g_n^{\bt}$ are normalization
constants.
They satisfy skew-orthonormality relations:

\begin{eqnarray}
\label{skew-norm} \int_{\mathbb
R}{\phi}^{(1)}_n(x)\psi_m^{(1)}(x)dx=Z_{n,m},\qquad
\int_{\mathbb
R}{\phi}^{(4)}_n(x)\psi_m^{(4)}(x)dx=\frac{Z_{n,m}}{2},
\qquad
Z = \left(\begin{array}{cc}
0 & 1     \\
-1 & 0         \\
\end{array}\right)\dotplus\ldots,
\qquad n,m\in\mathbb N.
\end{eqnarray}


Using these polynomials, we study the corresponding random matrix model:
\begin{equation}
P_{\beta,N}(H)dH=\frac{1}{{\cal Z}_{\beta N}}\exp[-[2{\rm
Tr}V(H)]]dH,\qquad \beta=1,4,
\end{equation}
where the matrix function $V(H)$ is a double well quartic polynomial of $H$ and
\begin{equation}
{\cal Z}_{\beta N}:= \int_{H\in M^{\bt}_{2N}}\exp[-[2{\rm
Tr}u(H)]]dH=(2N)!\prod_{j=0}^{2N-1}g_{j}^{\bt}.
\end{equation}
Here, $M^{\bt}_{2N}$ is a set of all $2N\times 2N$ real symmetric
($\beta=1$) and quaternion real self dual ($\beta=4$) matrices. $dH$
is the standard Haar measure.

To  study statistical properties of such matrix models,
we need to study certain kernel functions \cite{mehta1}:

\begin{eqnarray}
\nonumber \label{s} S^{(\beta)}_{2N}(x,y) := \sum_{j,k=0}^{2N-1}
Z_{j,k}\phi^{\bt}_{j}(x)\psi^{\bt}_{k}(y), \qquad
S^{(\beta)}_{2N}(y,x) = {S^{\dagger}}^{\bt}_{2N}(x,y).
\end{eqnarray}
For quartic potential, $\beta=1$, this is given by \cite{ghosh3}
\begin{eqnarray}
\nonumber \label{sn1}
(x-y)S^{(1)}_{2N}(x,y) &=&
 R^{(1)}_{2N-4,2N}[\psi^{(1)}_{2N-3}(x)\psi^{(1)}_{2N}(y)-(x\leftrightarrow y)]+
R^{(1)}_{2N-2,2N+2}[\psi^{(1)}_{2N-1}(x)\psi^{(1)}_{2N+2}(y)-(x\leftrightarrow y)]\\
\nonumber
&-& R^{(1)}_{2N-3,2N+1}[\psi^{(1)}_{2N-4}(x)\psi^{(1)}_{2N+1}(y)-(x\leftrightarrow y)]
-R^{(1)}_{2N-1,2N+3}[\psi^{(1)}_{2N-2}(x)\psi^{(1)}_{2N+3}(y)-(x\leftrightarrow y)]\\
\nonumber
&+& R^{(1)}_{2N-2,2N}[\psi^{(1)}_{2N-1}(x)\psi^{(1)}_{2N}(y)-(x\leftrightarrow y)]
 -  yP^{(1)}_{2N-3,2N}[\psi^{(1)}_{2N-4}(y)\psi^{(1)}_{2N}(x)-(x\leftrightarrow y)]\\
\nonumber
&-& [R^{(1)}_{2N-1,2N+1}[\psi^{(1)}_{2N-2}(x)\psi^{(1)}_{2N+1}(y)-(x\leftrightarrow y)]
 -  yP^{(1)}_{2N-1,2N+2}[\psi^{(1)}_{2N-2}(x)\psi^{(1)}_{2N+2}(y)-(x\leftrightarrow y)]]\\
&-& yP^{(1)}_{2N-2,2N+1}[\psi^{(1)}_{2N-1}(x)\psi^{(1)}_{2N+1}(y)-(x\leftrightarrow y)]
+ yP^{(1)}_{2N-1,2N}[\psi^{(1)}_{2N-2}(x)\psi^{(1)}_{2N}(y)-(x\leftrightarrow y)]
\end{eqnarray}
where the recursion coefficients $P^{(1)}_{j,k}$ and $R^{(1)}_{j,k}$ are defined as:
\begin{eqnarray}
\label{recursion-14}
\phi^{(1)}_{j}(x):=\sum_{k} P^{(1)}_{j,k}\psi^{(1)}_{k}(x);\qquad
x\phi^{(1)}_{j}(x):=\sum_{k} R^{(1)}_{j,k}\psi^{(1)}_{k}(x).
\end{eqnarray}
For $\beta=4$, $\psi^{(1)}_{j}(x)$ and $\phi^{(1)}_{j}(x)$
are replaced by $\phi^{(4)}_{j}(x)$ and $\psi^{(4)}_{j}(x)$ respectively \cite{ghosh3}.

To study asymptotic behavior of these skew-orthogonal functions
and their recursion coefficients $P^{\bt}_{j,k}$ and  $R^{\bt}_{j,k}$,
we expand $\phi^{(4)}_{m}(x)$ and $\psi^{(1)}_{m}(x)$,
$m\geq 1$ in a
suitable basis such that
their derivatives exist and
$\phi^{\bt}_{m}(x)$ are polynomials of order $m$.
We obtain recursion relations between sop and the corresponding op.
Solving the recursion relations and using asymptotic properties of op \cite{bleher}, we derive asymptotics
of sop. Finally, we apply them in the GCD formula (\ref{sn1}) to study the corresponding matrix models.


\subsection{SOP and Orthogonal Ensemble}

For sop with quartic weight, we expand $\psi^{(1)}_{m}(x)=\gamma^{(m)}_{m}\phi^{(2)}_{m-3}(x)+\sum^{m-1}\gamma^{(m)}_{k}\psi^{(1)}_{k}(x)$ such that
$[\psi_{m}^{(1)}]'(x)=\phi_{m}^{(1)}(x)$ is a polynomial of order $m$. Using skew-orthonormality
(\ref{skew-norm}), we get recursion relations:

\begin{eqnarray}
\label{psiodd1}
\psi^{(1)}_{2m+1}(x) &=& \gamma^{(2m+1)}_{2m+1}\phi^{(2)}_{2m-2}(x)+\gamma^{(2m+1)}_{2m-1}\psi^{(1)}_{2m-1}(x),
\qquad  m > 1,\\
\psi^{(1)}_{2m}(x) &=& \gamma^{(2m)}_{2m}\phi^{(2)}_{2m-3}(x)+\gamma^{(2m)}_{2m-2}\psi^{(1)}_{2m-2}(x)
+\gamma^{(2m)}_{2m-4}\psi^{(1)}_{2m-4}(x), \qquad  m > 1,\\
\phi^{(1)}_{2m+1}(x) &=& \gamma^{(2m+1)}_{2m+1}\left[P^{(2)}_{2m-2,2m+1}\phi^{(2)}_{2m+1}(x)+\dots
P^{(2)}_{2m-2,2m-5}\phi^{(2)}_{2m-5}(x)\right]
+\gamma^{(2m+1)}_{2m-1}\phi^{(1)}_{2m-1}(x),\\
\label{phieven1}
\phi^{(1)}_{2m}(x) &=& \gamma^{(2m)}_{2m}\left[P^{(2)}_{2m-3,2m}\phi^{(2)}_{2m}(x)+\dots
P^{(2)}_{2m-3,2m-6}\phi^{(2)}_{2m-6}(x)\right]
+\gamma^{(2m)}_{2m-2}\phi^{(1)}_{2m-2}(x)
+\gamma^{(2m)}_{2m-4}\phi^{(1)}_{2m-4}(x),
\end{eqnarray}
Here
\begin{equation}
\label{phij2}
\phi^{(2)}_{j}(x)=\frac{(x^j+\ldots)}{\sqrt{h_{j}}}\exp(-NV(x)),\qquad \int \phi^{(2)}_{j}(x)\phi^{(2)}_{k}(x)dx=\delta_{j,k}.
\end{equation}
are orthogonal polynomials. For even weight, with $R_{m}=(h_{m}/h_{m-1})$, $R_{0}=0$, op satisfy \cite{bleher}:
\begin{eqnarray}
\label{twotermrecursion}
&& x\phi^{(2)}_{j}(x) = \sqrt{R_{j+1}}\phi^{(2)}_{j+1}(x)+\sqrt{R_{j}}\phi^{(2)}_{j-1}(x)
;\qquad
\frac{d}{dx}\phi^{(2)}_{j}(x) = \sum_{k=j-3}^{j+3}P^{(2)}_{j,k}\phi^{(2)}_{k}(x)\\
\nonumber
&& P^{(2)}_{j,j+3} =  -Ng \sqrt{R_{j+1}R_{j+2}R_{j+3}},\qquad
P^{(2)}_{j,j+1}= -\frac{(j+1)}{2R_{j+1}^{1/2}},\qquad
P^{(2)}_{j,j-1} = \frac{j}{2R_{j}^{1/2}},\qquad
P^{(2)}_{j,j-3}= Ng\sqrt{R_{j-2}R_{j-1}R_{j}}.
\end{eqnarray}
Furthermore, using $\phi^{(2)}_{2m-1}(0)=0$ in (\ref{twotermrecursion}), we get:
\begin{eqnarray}
\label{phi2m0}
\phi^{(2)}_{2m}(0)={(-1)}^{m}\sqrt{\frac{R_{2m-1}\ldots R_{1}}{R_{2m}\ldots R_{2}}}\phi^{(2)}_{0}(0).
\end{eqnarray}

Using skew-orthogonality relations ($(\phi^{(1)}_{2m-2}(x),\psi^{(1)}_{2m+1}(x))=0$),
($(\phi^{(1)}_{2m-3}(x),\psi^{(1)}_{2m}(x))=0$) and ($(\phi^{(1)}_{2m-1}(x),\psi^{(1)}_{2m}(x))=0$)
in (\ref{psiodd1}) - (\ref{phieven1}), we get:
\begin{eqnarray}
\nonumber
\gamma^{(2m+1)}_{2m-1} &=& \gamma^{(2m+1)}_{2m+1}\gamma^{(2m-2)}_{2m-2} P^{(2)}_{2m-5,2m-2}; \qquad
\gamma^{(2m)}_{2m-4} = -\gamma^{(2m)}_{2m}\gamma^{(2m-3)}_{2m-3}P^{(2)}_{2m-6,2m-3}\\
\label{gammas}
\gamma^{(2m)}_{2m-2} &=& \gamma^{(2m)}_{2m}
\left[\gamma^{(2m-1)}_{2m-1}P^{(2)}_{2m-4,2m-3}
+\gamma^{(2m-1)}_{2m-3}\gamma^{(2m-3)}_{2m-3}P^{(2)}_{2m-6,2m-3}\right]
\end{eqnarray}
respectively.
We choose $\gamma^{(2m+1)}_{2m+1}=1$ and $\gamma^{(2m+1)}_{2m-1}=-\sqrt{R_{2m-2}/R_{2m-3}}$.
$\gamma^{(2m-2)}_{2m-2}$, $\gamma^{(2m)}_{2m-2}$ and $\gamma^{(2m)}_{2m-4}$ can be calculated
from (\ref{twotermrecursion}) and (\ref{gammas}).
Here, we note that choice of $\gamma^{(2m+1)}_{2m+1}$ and $\gamma^{(2m+1)}_{2m-1}$ is such that we can use
properties of orthogonal polynomials to obtain $\psi^{(1)}_{2m+1}(x)$ from (\ref{psiodd1}). However, this makes
the sop non-monic.

Using (\ref{twotermrecursion}) and (\ref{gammas}), we solve (\ref{psiodd1}).
$\psi^{(1)}_{2m+1}(x)$ while $\phi^{(1)}_{2m}(x)$
is obtained using skew-orthonormalization (\ref{skew-norm}):

\begin{eqnarray}
\label{psi-1-finite}
\psi^{(1)}_{2m+1}(x)  =  \sqrt{R_{2m-1}}\left[\frac{\phi^{(2)}_{2m-1}(x)}{x}\right],\qquad
\phi^{(1)}_{2m}(x)= \frac{x\phi^{(2)}_{2m-1}(x)}{\sqrt{R_{2m-1}}},\qquad m>1.
\end{eqnarray}
$\phi^{(1)}_{2m+1}(x)$ and $\psi^{(1)}_{2m}(x)$ can be obtained by taking derivative of $\psi^{(1)}_{2m+1}(x)$
and using skew-orthonormality respectively.

Now, we derive asymptotics for sop using results for op \cite{bleher}.
In the limit $m,N\rightarrow\infty$, we assume $\epsilon >0$, such that
$\epsilon<\lambda(=m/N) <\lambda_{cr}(=t^2/4g)$.
This ensures that $\phi^{\bt}_{m}(x)$ and $\psi^{\bt}_{m}(x)$
are concentrated on two intervals $[-x_{2},-x_{1}]$ and $[x_{1},x_{2}]$
and exponentially small outside.
Using results from \cite{bleher} for large $m$,
\begin{eqnarray}
\label{phi2-asy}
&& \phi^{(2)}_{m}(x)=\frac{2C_{m}\sqrt{x}}{\sqrt{\sin\theta}}[\cos(f_{m}(\theta))+O(N^{-1})],\qquad
f_{m}(\theta)=\left(\frac{m+1/2}{2}\right)
\left(\frac{\sin(2\theta)}{2}-\theta\right)-{(-1)}^{m}\frac{\chi}{4}+\frac{\pi}{4},
\\
\label{r2-asy}
&& R_{2m+1},R_{2m} = R,L+O(N^{-2})=\frac{-t\pm\sqrt{t^2-4\lambda g}}{2g},\qquad
{\rm with}\quad \lambda=gRL,\qquad t+g(R+L)=0.
\end{eqnarray}
Thus we have for $x_{1,2}=\sqrt{(-t\mp 2\sqrt{\lambda g})/g}$,
in the range $x_1+\delta<x<x_2-\delta$, $\delta > 0$ and $m >1$,
\begin{eqnarray}
\nonumber
&& \psi^{(1)}_{2m+1}(x)  \simeq
2C_{2m-1}\sqrt{\frac{{R}}{{x\sin\theta}}}\left[\cos\left(f_{2m-1}(\theta)\right)
+O\left(\frac{1}{N}\right)\right],\qquad
\psi^{(1)}_{2m}(x)  \simeq   \frac{C_{2m-1}}{N{\lambda}}
\sqrt{\frac{Lx}{\sin^{3}\theta}}
\left[\sin\left(f_{2m-1}(\theta)\right)
+O\left(\frac{1}{N}\right)\right],\\
\nonumber
\label{phi1odd}
&& \phi^{(1)}_{2m+1}(x)  \simeq
-4N\lambda C_{2m-1}\sqrt{\frac{x\sin\theta}{L}}
\left[\sin\left(f_{2m-1}(\theta)\right)
+O\left(\frac{1}{N}\right)\right],
 \phi^{(1)}_{2m}(x) \simeq   2C_{2m-1}\sqrt{\frac{x^3}{R\sin\theta}}
\left[\cos\left(f_{2m-1}(\theta)\right)
+O\left(\frac{1}{N}\right)\right],
\end{eqnarray}
where
\begin{eqnarray}
\label{asymp1}
&& q=\cos \theta = \frac{gx^2+t}{2\sqrt{{\lambda}'g}},\qquad \cos\chi=\frac{2\sqrt{{\lambda}'g}-tq}{{2\sqrt{{\lambda}'g}}q-t},\qquad \lambda=\frac{m}{N},
\qquad C_{m} = \frac{1}{2\sqrt{\pi}}{\left(\frac{g}{\lambda}\right)}^{1/4}(1+O(N^{-1})).
\end{eqnarray}
Here, we note, that for a given $x$, $\theta$ varies with $m$. We have chosen
$\theta\equiv\theta_{2m-1}$ such that
\begin{eqnarray}
\label{pmk}
\psi^{(1)}_{2m+1\pm k}(x) \simeq
2C_{2m-1}\sqrt{\frac{{R}}{{x\sin\theta}}}\left[\cos \left(f_{2m-1}(\theta)\mp (k\theta)/2\right)
+O(1/N)\right],\qquad m>>k,
\end{eqnarray}
and so on for $\psi^{(1)}_{2m\pm k}(x)$, $\phi^{(1)}_{2m+1\pm k}(x)$, $\phi^{(1)}_{2m\pm k}(x)$.

For the recursion coefficients (\ref{recursion-14}), we simply read off from
(\ref{psiodd1}) - (\ref{phieven1}). We get for large $m$:

\begin{eqnarray}
\nonumber
P^{(1)}_{2m,2m+3}=\sqrt{\frac{L}{R}};\qquad
P^{(1)}_{2m+1,2m+4}=R a^2;\qquad
P^{(1)}_{2m,2m+1}=-\frac{t}{gR};\qquad
P^{(1)}_{2m+1,2m+2}=0\\
\label{recursion1}
R^{(1)}_{2m,2m+4}=R^{(1)}_{2m+1,2m+5}=-N\lambda ; \qquad
R^{(1)}_{2m,2m+2}=R^{(1)}_{2m+1,2m+3}=Nt\sqrt{RL},
\end{eqnarray}
where $a=-N\sqrt{g\lambda}$.

To calculate $S_{2N}^{(1)}(x,y)$, we use (\ref{asymp1}) and (\ref{pmk})  in (\ref{sn1}). The first four terms
(modulo $O(N^{-1})$) in (\ref{sn1}) give $-(2\cos 2\theta\sin(\alpha_{2N-1}))/(\pi\sin^2\theta)$, the next four terms give $(2\cos^2\theta\sin(\alpha_{2N-1}))/(\pi\sin^2\theta)$
while the last two terms give $(\sin(\alpha_{2N-1}))/\pi$, where $\alpha_{j}=\Delta\theta(\partial f_{j}(\theta))/\partial\theta)$. Combining all, we get for $x=y+\Delta y$

\begin{eqnarray}
 (x-y)S_{2N}^{(1)}(x,y) = \frac{\sin(\alpha_{2N-1})}{\pi}+O(N^{-1})\qquad\Rrightarrow\qquad
 S_{2N}^{(1)}(x,y)  = \frac{\sin\left[\Delta y 2N\sqrt{g} y\sqrt{1-q^2}\right]}{\pi\Delta y}+O(N^{-1}).
 \end{eqnarray}
To obtain the level-density, we take the limit:
\begin{eqnarray}
\lim_{\Delta y\rightarrow 0}\frac{1}{2N}S_{2N}^{(1)}(x,y)=\frac{1}{2N}S_{2N}^{(1)}(y,y)  = \frac{\sqrt{g}}{\pi}|y|\sqrt{1-q^2}+O(N^{-1})\qquad
 = \frac{|y|}{\pi}\sqrt{g-{\left(\frac{gy^2+t}{2}\right)}^2}+O(N^{-1}),
 \end{eqnarray}
while $S_{2N}^{(1)}(x,y)/S_{2N}^{(1)}(y,y)$ gives the sine-kernel:
\begin{eqnarray}
\frac{S_{2N}^{(1)}(x,y)}{S_{2N}^{(1)}(y,y)}  = \frac{\sin \pi r}{\pi r},\qquad r=\Delta y S_{2N}^{(1)}(y,y).
 \end{eqnarray}

\section{SOP and Symplecic Ensemble}

For sop with quartic weight, we expand $\phi^{(4)}_{m}(x)=\gamma^{(m)}_{m}\phi^{(2)}_{m}(x)+\sum^{m-1}\gamma^{(m)}_{k}\phi^{(4)}_{k}(x)$. Using skew-orthonormality (\ref{skew-norm}), we get recursion relations:

\begin{eqnarray}
\label{phiodd4}
\phi^{(4)}_{2m+1}(x) &=& \gamma^{(2m+1)}_{2m+1}\phi^{(2)}_{2m+1}(x)+\gamma^{(2m+1)}_{2m-1}\phi^{(4)}_{2m-1}(x)\\
\phi^{(4)}_{2m}(x) &=& \gamma^{(2m)}_{2m}\phi^{(2)}_{2m}(x)+\gamma^{(2m)}_{2m-2}\phi^{(4)}_{2m-2}(x)
+\gamma^{(2m)}_{2m-4}\phi^{(4)}_{2m-4}(x)\\
\psi^{(4)}_{2m+1}(x) &=& \gamma^{(2m+1)}_{2m+1}\left[P^{(2)}_{2m+1,2m+4}\phi^{(2)}_{2m+4}(x)+\dots
P^{(2)}_{2m+1,2m-2}\phi^{(2)}_{2m-2}(x)\right]
+\gamma^{(2m+1)}_{2m-1}\psi^{(4)}_{2m-1}(x)\\
\label{psieven4}
\psi^{(4)}_{2m}(x) &=& \gamma^{(2m)}_{2m}\left[P^{(2)}_{2m,2m+3}\phi^{(2)}_{2m+3}(x)+\dots
P^{(2)}_{2m,2m-3}\phi^{(2)}_{2m-3}(x)\right]
+\gamma^{(2m)}_{2m-2}\psi^{(4)}_{2m-2}(x)
+\gamma^{(2m)}_{2m-4}\psi^{(4)}_{2m-4}(x),
\end{eqnarray}
where $\phi^{(2)}_{j}(x)$ are op defined in (\ref{phij2}) and (\ref{twotermrecursion}).

Using skew-orthogonality relations ($(\psi^{(4)}_{2m-2}(x),\phi^{(4)}_{2m+1}(x))=0$),
($(\psi^{(4)}_{2m-3}(x),\phi^{(4)}_{2m}(x))=0$) and ($(\psi^{(4)}_{2m-1}(x),\phi^{(4)}_{2m}(x))=0$)
in (\ref{phiodd4}) - (\ref{psieven4}), we get:
\begin{eqnarray}
\nonumber
\gamma^{(2m+1)}_{2m-1} &=& 2\gamma^{(2m+1)}_{2m+1}\gamma^{(2m-2)}_{2m-2} P^{(2)}_{2m-2,2m+1}; \qquad
\gamma^{(2m)}_{2m-4} = -2\gamma^{(2m)}_{2m}\gamma^{(2m-3)}_{2m-3}P^{(2)}_{2m-3,2m}\\
\label{gammas4}
\gamma^{(2m)}_{2m-2} &=& 2\gamma^{(2m)}_{2m}
\left[\gamma^{(2m-1)}_{2m-1}P^{(2)}_{2m-1,2m}
+\gamma^{(2m-1)}_{2m-3}\gamma^{(2m-3)}_{2m-3}P^{(2)}_{2m-3,2m}\right],
\end{eqnarray}
respectively. We choose $\gamma^{(2m+1)}_{2m+1}=1/\sqrt{2}$  and
$\gamma^{(2m+1)}_{2m-1}=-\sqrt{R_{2m+1}/R_{2m}}$.
$\gamma^{(2m-2)}_{2m-2}$, $\gamma^{(2m)}_{2m-2}$ and $\gamma^{(2m)}_{2m-4}$ can be calculated
from (\ref{twotermrecursion}) and (\ref{gammas4}).
Choice of $\gamma^{(2m+1)}_{2m+1}$ and $\gamma^{(2m+1)}_{2m-1}$ is such that we can use
properties of orthogonal polynomials to solve (\ref{phiodd4}). However, this makes
the sop non-monic.

Using (\ref{twotermrecursion}), (\ref{phi2m0}) and (\ref{gammas4}) we solve (\ref{phiodd4}).
 $\psi^{(4)}_{2m}(x)$
is obtained using skew-orthonormalization (\ref{skew-norm}):

\begin{eqnarray}
\label{phi-4-finite}
\phi^{(4)}_{2m+1}(x)=
\frac{\sqrt{R_{2m+2}}}{x\sqrt{2}}\left[\phi^{(2)}_{2m+2}(x)-\phi^{(2)}_{2m+2}(0)\exp[-NV(x)]\right],
\qquad
\psi^{(4)}_{2m}(x) = -\frac{x\phi^{(2)}_{2m+2}(x)}{\sqrt{2R_{2m+2}}}.
\end{eqnarray}
$\phi^{(4)}_{2m}(x)$ and $\psi^{(4)}_{2m+1}(x)$ can be derived by integrating and differentiating
$\psi^{(4)}_{2m}(x)$ and $\phi^{(4)}_{2m+1}(x)$ respectively.

With $m,N\rightarrow\infty$ and in the range $x_1+\delta<x<x_2-\delta$, $\delta > 0$, and neglecting the
$m$ independent term, we get
\begin{eqnarray}
\nonumber
\phi^{(4)}_{2m+1}(x) \simeq   C
\sqrt{\frac{2L}{x\sin\theta}}\left[\cos\left[f_{2m+2}(\theta)\right]
+O(1/N)\right],
\qquad
\phi^{(4)}_{2m}(x)  \simeq   -\frac{C}{N{\lambda}'}
\sqrt{\frac{Rx}{2\sin^{3}\theta}}
\left[\sin\left[f_{2m+2}(\theta)\right]
+O(1/N)\right],\\
\nonumber
\psi^{(4)}_{2m+1}(x)  \simeq
-\left[\frac{4NC{\lambda}'}{\sqrt{2R}}\right]\sqrt{x\sin\theta}
\left[\sin\left[f_{2m+2}(\theta)\right]
+O(1/N)\right],\qquad
\psi^{(4)}_{2m}(x)  \simeq  -C\sqrt{\frac{2x^3}{L\sin\theta}}
\left[\cos\left[f_{2m+2}(\theta)\right]
+O(1/N)\right],\\
\label{asymp4}
\end{eqnarray}
where $f_{m}(\theta)$, $C_m\equiv C$, $\chi$, $\lambda$ and $\theta$ are defined in (\ref{phi2-asy}) and
(\ref{asymp1}).
Here, we note, that for a given $x$, $\theta$ varies with $m$. We have chosen
$\theta\equiv\theta_{2m+2}$ such that
\begin{eqnarray}
\label{pmk4}
\phi^{(4)}_{2m+1\pm k}(x) \simeq   C
\sqrt{\frac{2L}{x\sin\theta}}\left[\cos\left[f_{2m+2}(\theta)\mp\frac{k\theta}{2}\right]
+O(1/N)\right],\qquad m>>k,
\end{eqnarray}
and so on for $\phi^{(4)}_{2m\pm k}(x)$, $\psi^{(4)}_{2m+1\pm k}(x)$, $\psi^{(4)}_{2m\pm k}(x)$.


For the recursion coefficients (\ref{recursion-14}), we simply read off from
(\ref{phiodd4}) - (\ref{psieven4}) and use (\ref{twotermrecursion}). For large $m$ we have:

\begin{eqnarray}
\nonumber
P^{(4)}_{2m,2m+3}=-\sqrt{\frac{R}{L}},\qquad
P^{(4)}_{2m+1,2m+4}=-{a}^2L,\qquad
P^{(4)}_{2m,2m+1}=\frac{t}{gL},\qquad
P^{(4)}_{2m+1,2m+2}=0,\\
\label{recursion4}
R^{(4)}_{2m,2m+4}=R^{(1)}_{2m+1,2m+5}=-N\lambda ,\qquad
R^{(4)}_{2m,2m+2}=R^{(1)}_{2m+1,2m+3}=Nt\sqrt{RL}.
\end{eqnarray}

To calculate $S_{2N}^{(4)}(x,y)$, we use (\ref{asymp4}), (\ref{pmk4}) and (\ref{recursion4}) in (\ref{sn1})
for $\beta=4$. The first four terms (modulo $O(N^{-1})$) in (\ref{sn1}) for $\beta=4$ give $-(\cos 2\theta\sin(\alpha_{2N+2}))/(\pi\sin^2\theta)$, the next four terms give $(\cos^2\theta\sin(\alpha_{2N+2}))/(\pi\sin^2\theta)$
while the last two terms give $(\sin(\alpha_{2N+2}))/\pi$. Combining all, we get for $x=y+\Delta y$,
\begin{eqnarray}
 (y-x)S_{2N}^{(4)}(x,y) = -\frac{\sin(\alpha_{2N+2})}{2\pi}+O(N^{-1})\qquad \Rrightarrow\qquad
 S_{2N}^{(4)}(x,y)  = \frac{\sin\left[\Delta y 2N\sqrt{g} y\sqrt{1-q^2}\right]}{2\pi\Delta y}+O(N^{-1})
 \end{eqnarray}
For level-density, we take the limit
\begin{eqnarray}
\lim_{\Delta y\rightarrow 0}\frac{1}{N}S_{2N}^{(4)}(x,y)=\frac{1}{N}S_{2N}^{(4)}(y,y)=\frac{\sqrt{g}}{\pi}|y|\sqrt{1-q^2}= \frac{|y|}{\pi}\sqrt{g-{\left(\frac{gy^2+t}{2}\right)}^2}+O(N^{-1})
 \end{eqnarray}
such that we get the ``universal'' sine-kernel
\begin{eqnarray}
\frac{S_{2N}^{(4)}(x,y)}{S_{2N}^{(4)}(y,y)}  = \frac{\sin 2\pi r}{2\pi r},\qquad r=\Delta y S_{2N}^{(4)}(y,y).
 \end{eqnarray}

\section{Conclusion}

The key achievement of this paper is the
derivation of Eqs. (\ref{psi-1-finite}) and (\ref{phi-4-finite}).
This enables us to use asymptotic results of op \cite{bleher} to derive
bulk asymptotics of sop with quartic weight \cite{stoj1,stoj1a,stoj2}. 
Simultaneously (\ref{psiodd1})-(\ref{phieven1}) 
and (\ref{phiodd4})-(\ref{psieven4}) gives us
the recursion coefficients $P^{\bt}_{j,k}$, $R^{\bt}_{j,k}$. These results for sop are applied
in the GCD formula \cite{ghosh3} to study quartic orthogonal and quartic symplectic ensembles of
random matrices in the bulk.
We note that asymptotics for sop away from the bulk can be trivially obtained 
from (\ref{psi-1-finite}) and (\ref{phi-4-finite}) 
since the corresponding results for op are already known.

\end{document}